# Origin of Strong Two-Magnon Scattering in Heavy Metal/Ferromagnet/Oxide Heterostructures

Lijun Zhu,[1]* Lujun Zhu,[2] D. C. Ralph,[1,3] and R. A. Buhrman[1]

1. Cornell University, Ithaca, New York 14850, USA
2. College of Physics and Information Technology, Shaanxi Normal University, Xi'an 710062, China
3. Kavli Institute at Cornell, Ithaca, New York 14853, USA

* lz442@cornell.edu

We experimentally investigate the origin of the two-magnon scattering (TMS) in heavy-metal (HM)/ferromagnet (FM)/oxide heterostructures (FM = Co, $Ni_{81}Fe_{19}$, or $Fe_{60}Co_{20}B_{20}$) by varying the materials located above and below the FM layers. We show that strong TMS in HM/FM/oxide systems arises primarily at the HM/FM interface and increases with the strength of interfacial spin-orbit coupling and magnetic roughness at this interface. TMS at the FM/oxide interface is relatively weak, even in systems where spin-orbit coupling at this interface generates strong interfacial magnetic anisotropy. We also suggest that the spin-current-induced excitation of non-uniform short-wavelength magnon at the HM/FM interface may function as a mechanism of spin memory loss for the spin-orbit torque exerted on the uniform mode.

## I. Introduction

The magnetic damping ($\alpha$) of magnetic thin-film systems is a key parameter in the determination of the relaxation time of magnetization dynamics [1], the propagation distance of spin waves [2], the speed of antidamping torque switching of a macrospin [3], the velocity of current-induced skyrmion motion [4], and the energy efficiency of spin-torque magnetic memories [5,6], oscillators [7] and logic [8]. For in-plane magnetized heavy metal (HM)/ferromagnet (FM)/oxide heterostructures, the variation of $\alpha$ with the FM thickness ($t_{FM}$) has been widely used to determine the effective spin-mixing conductance ($G_{eff}^{\uparrow\downarrow}$) of HM/FM interfaces for spin transport analysis based on the widespread assumption that spin pumping is the dominant source of $\alpha$ [9-16]. Recently, we have demonstrated that this assumption fails badly for most sputter-deposited in-plane magnetized HM/FM/oxide heterostructures in nanometer-scale $t_{FM}$ regime of interest for spintronic devices [17], leading to far-reaching consequences including considerable overestimation of the true $G_{eff}^{\uparrow\downarrow}$ and large errors in any analysis that relies on the value of $G_{eff}^{\uparrow\downarrow}$. Instead, we found that $\alpha$ in these sputter-deposited multilayers is generally dominated by two-magnon scattering (TMS)[17], in which a uniform magnon of a precessional macrospin scatters into a degenerate non-uniform short-wavelength magnon induced by material imperfections [18-20].

In this work, we investigate in detail the origin of TMS in HM/FM/oxide devices as well as the potential influence of non-uniform magnons on the efficiency of spin-orbit torques (SOTs). By varying the underlayer and the overlayer for different FM metals (e.g. Co, $Ni_{81}Fe_{19}$, and $Fe_{60}Co_{20}B_{20}$), we find that TMS in our HM/FM/oxide devices arises primarily from the interfacial spin-orbit coupling (ISOC) and magnetic roughness of the HM/FM interface. In contrast, TMS at the FM/oxide interface is much weaker even when strong ISOC at this interface generates large interfacial perpendicular magnetic anisotropy. We also suggest that, when the ISOC is strong at the HM/FM interface, the spin-current-induced excitation of non-uniform short-wavelength magnons at this interface may influence SOT experiments as a form of a spin memory loss (SML).

## II. Background

The magnetic damping of a HM/FM/oxide system can be understood as the sum of "intrinsic" damping of the FM layer [21] and interfacial damping including spin pumping and two-magnon scattering processes, i.e., $\alpha = \alpha_{int} + \alpha_{SP} + \alpha_{TMS}$ [17] or

$$\alpha = \alpha_{int} + (G_{eff}^{\uparrow\downarrow} + G_{SML})\frac{g\mu_B h}{4\pi M_s e^2} t_{FM}^{-1} + \beta_{TMS} t_{FM}^{-2}. \quad (1)$$

Here $g$ is the g-factor, $\mu_B$ the Bohr magnetron, $h$ the Planck's constant, and $M_s$ the saturation magnetization of the FM layer. The second term of Eq. (1) is the combined contribution ($\alpha_{SP}$) from spin pumping spin current that is absorbed in the HM layer [9-14] and at the HM/FM interface due to SML [16,22,23], which for convenience we parameterize as an "effective SML conductance" $G_{SML}$. The third term of Eq. (1), $\alpha_{TMS} = \beta_{TMS} t_{FM}^{-2}$, is the TMS damping arising from the combination of ISOC and magnetic roughness (e.g., variations of the thickness and/or the ISOC strength) [17,18-20]. The coefficient $\beta_{TMS} = C_{TMS}(2K_s^{ISOC}/M_s)^2$ [17,18], where $C_{TMS}$ is a parameter related to the density and the geometry of the scattering defects at the interfaces [19,20] and $K_s^{ISOC}$ is the interfacial magnetic energy density of the associated interface. In most HM/FM/oxide heterostructures with $t_{FM}$ only a few nm thick, $\alpha_{TMS} >> \alpha_{SP}$ [17].

## III. Sample configurations

As listed in detail in Table 1, the magnetic stacks studied for this work consist of six Co-based sample series (A1-A6), four $Ni_{81}Fe_{19}$-based sample series (B1-B4), and six $Fe_{60}Co_{20}B_{20}$-based sample series (C1-C6). For each series, different materials were deposited below or/and above the FM layer. All the stacks were deposited by DC/RF sputtering onto 4" oxidized silicon substrates. For most samples, the FM layer was prepared by oblique deposition to make a wedge, allowing devices with different FM thicknesses to be studied across a single wafer. Each FM wedge was 75 mm long. The thickness slopes are ≈ $6.9 \times 10^{-5}$ Å/μm for the Co wedges, $4.6 \times 10^{-5}$ Å/μm for the $Ni_{81}Fe_{19}$ wedges, and $(5.9\text{-}9.0) \times 10^{-5}$ Å/μm for the $Fe_{60}Co_{20}B_{20}$ wedges. Some samples were also made with constant ferromagnetic thicknesses by rotating the substrates during deposition, from which we verified that the oblique sputtering of the wedges did not affect the results (see below). A 1 nm Ta seed layer was deposited as part of the growth of all the sample series other than A1 and B1. All samples with a MgO layer were capped with 1.5 nm of Ta.

Sample series A1-A6, B1-B3, and C1 (Table 1) were patterned into magnetic strips (10×20 μm$^2$) via photolithography and ion milling to measure $\alpha$ and the effective demagnetization field ($4\pi M_{\text{eff}}$) and the SOT efficiencies using spin-torque ferromagnetic resonance (ST-FMR) [24,25] and also into Hall bars (5 × 65 μm$^2$) for determination of SOT efficiencies using "in-plane" harmonic Hall response technique [26-28]. $\alpha$ and $4\pi M_{\text{eff}}$ of sample series B4 and C2-C6 (Table 1) were obtained from unpatterned pieces using a flip-chip FMR with the magnetic pieces face-side down on a co-plane waveguide.

## IV. Source of two-magnon scattering in HM/FM/oxide structures

Using ST-FMR [24,25] and flip-chip FMR in the rf frequency ($f$) regime of 7-18 GHz, we measured $\alpha$ and $4\pi M_{\text{eff}}$ for each sample from the best fits of the FMR linewidth ($\Delta H$, half width at half maximum) and resonance field ($H_r$) to the relations [29]:

$$\Delta H = \Delta H_0 + 2\pi \alpha f / \gamma, \quad (2)$$
$$f = (\gamma/2\pi)\sqrt{H_r(H_r + 4\pi M_{\text{eff}})}, \quad (3)$$

where $\Delta H_0$ is the inhomogeneous broadening of the FMR linewidth and $\gamma$ is the gyromagnetic ratio. In Fig. 1(a), we plot the values of $\alpha$ for SiO$_2$/Co/Pt (series A1), Pt/Co/MgO (series A2) Pt/Hf/Co/MgO (series A3), and Ta/Co/MgO (series A4) as a function of $t_{\text{Co}}^{-1}$. We find that $\alpha$ for Ta/Co/MgO (A4) remains small and almost constant as a function of $t_{\text{Co}}^{-1}$, corresponding to $\alpha_{\text{int}} = 0.0126 \pm 0.0001$ for the Co layer, with negligible amounts of both TMS and spin pumping. From this, we can exclude both the Ta/Co and Co/MgO interfaces as strong sources of TMS. This result is consistent with previous measurements of weak damping enhancement at Ta/FM interfaces [14,15], and the weak ISOC of Ta/Co (see below).

In contrast, the values of $\alpha$ for sample series A1-A3 vary proportional to $t_{\text{Co}}^{-2}$, indicating that TMS is the dominant mechanism of magnetic damping in these heterostructures. From the best fits of the data in Fig. 1(a) to

$$\alpha = \alpha_{\text{int}} + \beta_{\text{TMS}} t_{\text{FM}}^{-2} \quad (4)$$

we obtain the values listed in Table 1 for $\beta_{\text{TMS}}$, which parameterizes the strength of TMS. $\beta_{\text{TMS}}$ is substantial in all three sample series and is almost one order of magnitude larger for SiO$_2$/Co/Pt (A1) than for Pt/Co/MgO (A2) and Pt/Hf/Co/MgO (A3). As expected [17,18], the strength of TMS is correlated with the strength of ISOC and magnetic roughness. We can determine the total interfacial magnetic anisotropy density ($K_s$, the sum from both interfaces of the FM) and the saturation magnetization $M_s$ using fits of $4\pi M_{\text{eff}}$ vs. $t_{\text{Co}}^{-1}$ (Fig. 1(b)) to the relation [29]

$$4\pi M_{\text{eff}} \approx 4\pi M_s + 2K_s/M_s t_{\text{Co}}. \quad (5)$$

For SiO$_2$/Co/Pt (A1) we find $K_s = 2.31 \pm 0.05$ erg/cm$^2$ and $M_s = 1417 \pm 30$ emu/cm$^3$, for Pt/Co/MgO (A2) $K_s = 1.78 \pm 0.01$ erg/cm$^2$ and $M_s = 1314 \pm 8$ emu/cm$^3$, for Pt/Hf/Co/MgO (A3) $K_s = 1.21 \pm 0.03$ erg/cm$^2$ and $M_s = 1352 \pm 12$ emu/cm$^3$, and for for Ta/Co/MgO (A4) $K_s = 0.34 \pm 0.04$ erg/cm$^2$ and $M_s = 1134 \pm 16$ emu/cm$^3$. If we assume that the small $K_s$ for Ta/Co/MgO (A4) is due mostly to the Co/MgO interface (i.e. $K_s^{\text{Co/MgO}} \approx 0.34$ erg/cm$^2$) and that $K_s$ is zero for the SiO$_2$ interface, we can estimate $K_s^{\text{ISOC}}$ for the individual HM/FM interfaces to be for Co/Pt (A1) $K_s^{\text{ISOC}} = 2.31 \pm 0.05$ erg/cm$^2$, for Pt/Co (A2) $K_s^{\text{ISOC}} = 1.44 \pm 0.01$ erg/cm$^2$, and for Pt/Hf/Co (A3) $K_s^{\text{ISOC}} = 0.86 \pm 0.03$ erg/cm$^2$. In Fig. 1(c) we plot $\beta_{\text{TMS}}$ for these four series of Co samples as a function of $(2K_s^{\text{ISOC}}/M_s)^2$. For the three samples (A2-A4) deposited with a Ta seed layer, we find an accurate linear scaling. This is consistent with the expectation for TMS [i.e. $\beta_{\text{TMS}} = C_{\text{TMS}}(2K_s^{\text{ISOC}}/M_s)^2$] and indicates a similar magnetic roughness of these HM/Co interfaces ($C_{\text{TMS}} \approx 0.08$ T$^2$). In contrast, $\beta_{\text{TMS}}$ for SiO$_2$/Co/Pt (A1) is 4-fold higher than extrapolated from the linear fit in Fig. 1(c), suggesting a considerable increase in magnetic roughness. Cross-sectional transmission electron microscopy (TEM) images of the samples are shown in Figs. 1(d)-1(f). In SiO$_2$/Co/Pt (A1, Fig. 1(d)), the Co layer, whose nominal "thickness was 2.3 nm, has a granular texture and is thus magnetically very rough. This granularity arises because Co has much higher surface energy than SiO$_2$, while the Pt grows coherently on the Co grains. In contrast, for Pt/Co/MgO (A2, Fig. 1(e)) the Co layer is atomically smooth at both interfaces and coherently follows the Pt lattice. When a 0.3 nm Hf layer is inserted at Pt/Co interface in sample series (A3, Fig. 1(f)), the Co layer still grows in a relatively smooth manner, while its coherent growth is substantially interrupted by the Hf insertion as indicated by the distortion of the lattice planes in the Co layer near the interface with Pt/Hf. In both Pt/Co/MgO (A2) and Pt/Hf/Co/MgO (A3), the Co/MgO interface is atomically sharp (Figs. 1(e) and 1(f)), consistent with a negligible TMS at the Co/MgO interfaces. The relatively strong TMS at the atomically smooth Pt/Co interfaces in Pt/Co/MgO (A2) and Pt/Hf/Co/MgO (A3) suggests that the observed TMS in those samples is due mainly to the ISOC variation induced by the polycrystalline texture (e.g., the different orientations and dimensions of crystalline grains) rather than a thickness-induced roughness. The much stronger TMS in SiO$_2$/Co/Pt (A1) may also have a contribution from the much larger roughness in those samples.

Our results for samples with Ni$_{81}$Fe$_{19}$ magnetic layers (sample series B1-B4 shown in Fig. 2) are similar to the Co-based samples. TMS at Ni$_{81}$Fe$_{19}$/MgO interfaces is minimal, see MgO/Ni$_{81}$Fe$_{19}$/MgO (B4) in Fig. 2(a). The presence of a Pt/Ni$_{81}$Fe$_{19}$ interface enhances TMS for Pt/Ni$_{81}$Fe$_{19}$/MgO (B2, B3), and the TMS is the largest in the sample series without a smoothing Ta seed layer, SiO$_2$/Ni$_{81}$Fe$_{19}$/Pt (B1). The value of $\beta_{\text{TMS}}$ is small for MgO/Ni$_{81}$Fe$_{19}$/MgO (B4) despite the fact that $(2K_s/M_s)^2$ is 1.5 times larger for this sample than for Pt/Ni$_{81}$Fe$_{19}$/MgO (B2, B3) (see Table 1 and Fig. 2(b)). Within the usual model of TMS [17,18], this suggests that the Ni$_{81}$Fe$_{19}$/MgO interfaces are magnetically smooth, with small values of $C_{\text{TMS}}$.

As we show in Fig. 3(a), TMS in the HM/Fe$_{60}$Co$_{20}$B$_{20}$/MgO sample series is weaker than that in Co or Ni$_{81}$Fe$_{19}$ samples, but nevertheless it is clearly measurable. The small values of $\beta_{TMS}$ for the Fe$_{60}$Co$_{20}$B$_{20}$ samples (e.g. 0.04 nm$^{-2}$ for Pt/Fe$_{60}$Co$_{20}$B$_{20}$/MgO and 0.09 nm$^{-2}$ for Ta/Fe$_{60}$Co$_{20}$B$_{20}$/MgO) are consistent with the relatively weak ISOC (Fig. 3(b)). We continue to find that $\alpha$ in the HM/Fe$_{60}$Co$_{20}$B$_{20}$/MgO systems is strongly dependent on the details of HM/Fe$_{60}$Co$_{20}$B$_{20}$ interface but insensitive to even strong ISOC at the Fe$_{60}$Co$_{20}$B$_{20}$/MgO interface. For instance, as we show in Figs. 4(a) and 4(b), $\alpha$ of Au$_{25}$Pt$_{75}$ 4/Fe$_{60}$Co$_{20}$B$_{20}$ 1.6/MgO (C4) is reduced markedly when a spacer bilayer of Pt 0.5/Co 0.25 is inserted at the Au$_{25}$Pt$_{75}$/Fe$_{60}$Co$_{20}$B$_{20}$ interface (sample C5) likely due to a reduction in ISOC (as indicated by an increased value of $4\pi M_{eff}$, Fig. 4(b)). However, $\alpha$ remains similar when $4\pi M_{eff}$ is reduced by up to 50% due to the insertion of a 0.1 nm Hf spacer at the Fe$_{60}$Co$_{20}$B$_{20}$/MgO interface (sample C6). The same is true after the samples were annealed at 450 °C for 1 hour (Fig. 4(c) and 4(d)). This is consistent with previous measurements of a substantial reduction of $\alpha$ for (Pt or Pt$_{85}$Hf$_{15}$)/Fe$_{60}$Co$_{20}$B$_{20}$/MgO by insertion of an ultrathin Hf spacer in between the HM and FM layers [30,31]. An earlier pump-probe magneto-optical Kerr effect experiment [14] also indicated that the increase of the in-plane damping with $t_{FM}^{-1}$ is approximately a factor of 2 faster for Ta 5/Co$_{40}$Fe$_{40}$B$_{20}$/Ta 5 (annealed at 250 °C) than for Ta 5/Fe$_{40}$Co$_{40}$B$_{20}$/MgO (annealed at 250 °C). These observations consistently reveal that TMS of HM/FeCoB/MgO arises primarily from the HM/FeCoB interface, while the FeCoB/MgO interface is magnetically smoother and contributes minimally to $\alpha$ despite the fact that it is the primary source of the total ISOC and the interfacial PMA. This is a technologically interesting observation because it indicates that HM/FeCoB/oxide devices can be tuned to have a low $\alpha$ and low $4\pi M_{eff}$ (high PMA) at the same time by separately reducing ISOC at the HM/FeCoB interface and enhancing ISOC at the FeCoB/oxide interface.

We do find significant TMS ($\beta_{TMS} \approx 0.2$ nm$^{-2}$) in MgO/Fe$_{60}$Co$_{20}$B$_{20}$/MgO (C3, Fig. 3(a)), suggesting a large magnetic roughness and an enhanced ISOC at the MgO/Fe$_{60}$Co$_{20}$B$_{20}$ interface. This is similar to the roughness from SiO$_2$/Co (A1, Fig. 1(a)), SiO$_2$/Ni$_{81}$Fe$_{19}$ (B1, Fig. 2(a)), and previous measurements of SiO$_2$/Ni$_{50}$Fe$_{50}$ [19]. The increased magnetic roughness is likely because the surface energy of the metallic FMs is higher that MgO and SiO$_2$.

### V. Insensitivity to oblique growth of the FM layer

It is well established that the strength of TMS can vary when the magnetic precession axis is oriented along different directions with respect to an anisotropic surface defect [18-20]. Oblique deposition of thin-film wedges has the potential to induce anisotropic tilting of crystalline grains as well as variations in film thickness [32-35]. Here we affirm that our analysis on TMS is not affected by the oblique depositions we used to make wedged samples.

We first patterned identical ST-FMR microstrips (20 × 30 μm$^2$) with different orientations ($\varphi = 0$°, ±30°, ±45°, ± 60°, and 90° relative to the wedge gradient (see Fig. 5(a)) for each fixed value of $t_{FM}$ from Pt 5.3/Co 1.8-6.6/MgO (sample A5). In all cases, ST-FMR was performed with an applied field oriented at a fixed angle of 45° from the microstrip axis. As shown in Figs. 5(b) and 5(c), ST-FMR measurements of both $\alpha$ and $4\pi M_{eff}$ are independent of $\varphi$ within our experimental sensitivity, indicating absence of any anisotropy due to the orientation of the magnetic precession axis relative to the wedge direction. In Fig. 5(c), we show that the $\alpha$ values of the wedged sample (A2) agree with those of sample A6 where the Co layers are grown uniformly with substrate rotation during growth. We also find no indication of any sensitivity to oblique deposition of $\alpha$ or $4\pi M_{eff}$ for the Ni$_{81}$Fe$_{19}$ layer in the Pt/Ni$_{81}$Fe$_{19}$ bilayers (sample B1 and B2). From these observations, we can safely conclude that the oblique deposition of the FM wedge is not an important source of the observed TMS in our HM/FM/oxide systems.

### VI. Influence of short-wavelength magnons on spin torque

It is known that magnons can efficiently inter-convert with spin currents and generates inverse spin Hall voltage [36,37] or SOTs [38]. It is, therefore, an important question as to how the non-uniform short-wavelength magnons at a HM/FM interface affects the SOT efficiency in the heterostructure. As schematically shown in Fig. 6(a), the non-uniform short-wavelength magnons might affect a SOT measurement via four possible processes: (i) they can be excited directly by the spin current from the HM layer and relax into the lattice; (ii) they can be excited by relaxation of the uniform magnon mode and then subsequently relax into the lattice; (iii) they can be excited by the spin current from the HM layer and then relax by transferring spin angular momentum to the uniform mode; (iv) they can be excited by the rf Oersted field and then relax by transferring angular momentum to the uniform mode. In the first process, the non-uniform magnons would behavior as a source of SML that reduces the efficiency of SOT; in the second process, the short-wavelength magnons would provide an additional channel for damping in the second process. Processes (iii) and (iv) would enhance the SOT efficiency for the uniform mode if the inter-conversion of the spin current is more efficient with the short-wave magnon than with uniform mode (e.g. as indicated at YIG/HM [36,37]).

To examine the possible effects of the non-uniform short-wavelength magnons on the SOT exerted on the uniform mode, we determined the dampinglike SOT efficiency for the Co/Pt (A1), Pt/Co (A2) and Pt/Hf/Co (A3) using both ST-FMR measurements [24] and harmonic response measurements [26,27]. For the ST-FMR determination, if we employ the standard macrospin analysis and assume a negligible spin pumping effect, we can define an effective FMR spin-torque efficiency $\xi_{FMR}$ [24] from the ratio of the symmetric (S) and anti-symmetric (A) components of the magnetoresistance response of the ST-FMR resonance. S is proportional to $H_{DL}$ and A is due to the sum of the Oersted field and the fieldlike SOT effective field. The dampinglike and fieldlike SOT efficiencies per applied electric field, $\xi_{DL}^{E}$ and $\xi_{FL}^E$, can then be obtained from the linear dependence of $\zeta_{FMR}^{-1}$ on $t_{FM}^{-1}$ when $\xi_{DL}^E$, $\xi_{FL}^E$, the HM resistivity ($\rho_{HM}$), and $M_s$ are approximately constant over the studied $t_{FM}$ regime [24],

$$\frac{1}{\xi_{FMR}} = \frac{1}{\xi_{DL}^E \rho_{HM}} \left(1 + \frac{\hbar}{e} \frac{\xi_{FL}^E \rho_{HM}}{\mu_0 M_s d} \frac{1}{t_{FM}}\right). \quad (6)$$

Here, $\rho_{xx}$ of the 4.5 nm Pt layer was 61 μΩ cm for Co/Pt, 35 μΩ cm for Pt/Co, 40 μΩ cm for Pt/Hf/Co. As plotted in Fig. 6(b), $\xi_{DL}^E$ was estimated from the ST-FMR measurement to be $(0.98 \pm 0.03) \times 10^5$ $\Omega^{-1}$ $m^{-1}$ for Co/Pt (A1), to $(1.62 \pm 0.04) \times 10^5$ $\Omega^{-1}$ $m^{-1}$ for Pt/Co (A2), and $(2.64 \pm 0.29) \times 10^5$ $\Omega^{-1}$ $m^{-1}$ for Pt/Hf 0.3/Co (A3).

For harmonic response measurement on Co 2.3/Pt 4.5 (A1), Pt 4.5/Co 2.5 (A2), and Pt 4.5/Hf 0.3/Co 2.3 (A3), the second harmonic Hall voltage response ($V_{2\omega}$) was measured as a function of in-plane orientation of magnetization ($\varphi$) under different fixed magnitudes of in-plane magnetic bias field ($H_{in}$) (1-3.5 kOe), under the excitation of a low-frequency sinusoidal electric field (61.5 kV/m, 1.327 kHz) on Hall bars. As described in more detail in Refs. [27,28], the $\cos\varphi$ dependent component ($V_a$) of $V_{2\omega}$ follows

$$V_a = -H_{DL}V_{AH}/2(H_{in}+ H_k) + V_{ANE}, \quad (7)$$

where $H_{DL}$ is the dampinglike SOT effective field, $V_{AH}$ the anomalous Hall voltage, $H_k$ the anisotropic field, and $V_{ANE}$ the anomalous Nernst voltage. Using the values of $H_{DL}$ given by the slope of the linear fits of the $V_a$ data to Eq. (7) (Fig. 6(d)), we determined $\xi_{DL}^E$ of the samples following $\xi_{DL}^E = (2e/\hbar)\mu_0 H_{DL} M_s t_{Co}/E$. As plotted in Fig. 6(d), $\xi_{DL}^E$ obtained from the harmonic response measurement increases from $(3.17 \pm 0.11) \times 10^5$ $\Omega^{-1}$ $m^{-1}$ for Co 2.3/Pt 4.5 (A1), to $(4.58 \pm 0.05) \times 10^5$ $\Omega^{-1}$ $m^{-1}$ for Pt 4.5/Co 2.5 (A2), and to $(5.69 \pm 0.87) \times 10^5$ $\Omega^{-1}$ $m^{-1}$ for Pt 4.5/Hf 0.3/Co 2.3 (A3).

The $\xi_{DL}^E$ values measured using either ST-FMR or harmonic response decreases with $K_s$ approximately in a linear manner, which together with our previous observation in the HM/Co bilayers annealed at different conditions [39] indicates a linear decrease in the spin transparency of the interface ($T_{int}$). This is because $\xi_{DL}^E = T_{int}\sigma_{SH}$ for a HM/FM bilayer and the spin Hall conductivity of the HM ($\sigma_{SH}$) is constant when $\rho_{xx}$ is constant [27,28]. $T_{int}$ for a SOT process should be given by

$$T_{int} = G_{HM/FM}^{\uparrow\downarrow}/(G_{HM/FM}^{\uparrow\downarrow} + G_{SML} + G_{HM}/2), \quad (8)$$

where $G_{HM/FM}^{\uparrow\downarrow}$ is the bare spin-mixing conductance of the interface, $G_{HM} = 1/\rho_{xx}\lambda_s$ and $\lambda_s$ are the spin conductance and the spin diffusion length of the HM. $G_{HM}$ should be constant within the Elliot-Yafet spin relaxation mechanism [40,41], while $G_{HM/FM}^{\uparrow\downarrow}$ and $G_{SML}$ can be modulated by changes at the interface [42,43]. The monotonic decrease of $T_{int}$ with $K_s$ should indicate an increase of $G_{SML}$ with $K_s$. This might be suggestive of the possibility that the non-uniform magnons are excited directly by spin current from the HM layer and relax into the lattice (the aforementioned processes (i)). The decrease of $T_{int}$ with $K_s$ is less likely to suggest a decrease in $G_{HM/FM}^{\uparrow\downarrow}$ here because previous studies have indicated that magnetic roughness (e.g. induced by diffusion) may increase $T_{int}$ via moderately enhancing $G_{HM/FM}^{\uparrow\downarrow}$ [42,43]. Finally, while processes (iii) and (iv) that should increase $\xi_{DL}^E$ are still possible, they seem to be a weaker effect than the SML process as indicated by the decrease of $T_{int}$ with enhancing $K_s$.

It is also an interesting observation that the values of $\xi_{DL}^E$ obtained from the standard ST-FMR analysis are more than a factor of 2 smaller than those obtained from the harmonic response measurements (Fig. 6(d)), with this ratio getting larger as ISOC becomes greater. However, this difference cannot be fully attributed to the excitation of the short-wavelength magnons because the difference still seems to exist at zero ISOC as indicated by the extrapolation of the data to zero $K_s$ (straight lines in Fig. 6(d)). We note that the ST-FMR measurements here are accompanied by significant spin pumping in the *thick* Co regime as indicated by the deviation of $\zeta_{FMR}^{-1}$ from the linear $t_{FM}^{-1}$ dependence (Fig. 6(c)), while the spin pumping seems negligible in the thin Co regime where we took data to determine $\xi_{DL}^E$ according to Eq. (6). It warrants future efforts to unveil the cause of the different values of $\xi_{DL}^E$, but that is beyond the scope of this work. Here, it worth mentioning that the "in-plane" harmonic Hall response measurements, if performed carefully [26,27], yield results of $\xi_{DL}^E$ that are consistent with those obtained from "out-of-plane" harmonic Hall response measurements [44,45] and antidamping SOT switching of in-plane magnetized 3-terminal magnetic tunnel junctions [46,47].

## VI. Conclusion

We have shown that the strong extrinsic damping in HM/FM/oxide systems arises dominantly from the TMS due to the coexistence of ISOC and magnetic roughness at the HM/FM interfaces, while is largely irrelevant to FM/oxide interfaces and the oblique growth of the FM layer. These results indicate that the energy efficiency of SOT-driven magnetic memories, oscillators, and logic devices, where HM/FM/oxide is the core ingredient, can be substantially improved by separately reducing the ISOC at the HM/FM interface while enhancing the ISOC at the FM/oxide interface through interface engineering. We also suggest that the short-wavelength magnons may be excited by spin current from the HM layers and subsequently relax into the lattice, and thus function as a source of SML in a SOT process. These results indicate that the ISOC and magnetic roughness at the HM/FM interfaces should be *minimized* in spin-torque memories and logic where high spin-torque efficiency and low damping are required to reduce the power consumption.

This work was supported in part by the Office of Naval Research (N00014-15-1-2449), in part by the NSF MRSEC program (DMR-1719875) through the Cornell Center for Materials Research, and in part by the NSF (ECCS-1542081) through use of the Cornell Nanofabrication Facility/National Nanotechnology Coordinated Infrastructure. The TEM measurements performed at Shaanxi Normal University were supported by the National Natural Science Foundation of China (Grant No. 51901121), the Science and Technology Program of Shaanxi Province (Grant No. 2019JQ-433), and the Fundamental Research Funds for the Central Universities (Grant No. GK201903024).

Table 1. Sample configurations, with layer sequences listed from the bottom ($Si/SiO_2$ substrate) to the top. Numbers are layer thicknesses in nm. Sample series A1-A6, B1-B4, and C1-C6 are heterostructures based on Co, $Ni_{81}Fe_{19}$, and $Fe_{60}Co_{20}B_{20}$, respectively. For the sample series grown with a wedged ferromagnetic layer, measurements as a function of ferromagnet thickness allow measurements of the coefficient of two-magnon scattering ($\beta_{TMS}$) and the ratio of interface magnetic anisotropy to the saturation magnetization ($2K_s/M_s$).

| # | Magnetic heterostructures | $\beta_{TMS}$ (nm$^2$) | $2K_s/M_s$ (T nm) | technique |
|---|---|---|---|---|
| A1 | $Si/SiO_2$/Co 1.8-6.6 (wedge)/Pt 4.5 | 2.73 ±0.42 | -3.48 ±0.10 | ST-FMR |
| A2 | $Si/SiO_2$/Ta 1/Pt 4.5/Co 1.8-6.6 (wedge)/MgO 2 | 0.41 ±0.04 | -2.69 ±0.03 | ST-FMR |
| A3 | $Si/SiO_2$/Ta 1/Pt 4.5/Hf 0.3/Co 1.8-6.6 (wedge)/MgO 2 | 0.32 ±0.01 | -1.72 ±0.08 | ST-FMR |
| A4 | $Si/SiO_2$/Ta 1/Co 1.8-6.6 (wedge)/MgO 2 | 0 | -0.74 ±0.01 | ST-FMR |
| A5 | $Si/SiO_2$/Ta 1/Pt 5.3/Co 1.8-6.6 (wedge)/MgO 2 | 1.38 ±0.10 | -1.85 ±0.05 | ST-FMR |
| A6 | $Si/SiO_2$/Ta 1/Pt 4.5/Co 1.8, 2, 3, 5/MgO 2 | - | - | ST-FMR |
| B1 | $Si/SiO_2$/$Ni_{81}Fe_{19}$ 1.8-5 (wedge)/Pt 4.5 | 0.41 ±0.01 | -1.80 ±0.07 | ST-FMR |
| B2 | $Si/SiO_2$/Ta 1/Pt 4.5/$Ni_{81}Fe_{19}$ 1.8, 2, 3/MgO 2 | - | - | ST-FMR |
| B3 | $Si/SiO_2$/Ta 1/Pt 4.5/$Ni_{81}Fe_{19}$ 1.8-5 (wedge)/MgO 2 | 0.30 ±0.02 | -1.04 ±0.05 | ST-FMR |
| B4 | $Si/SiO_2$/Ta 1/MgO 2/$Ni_{81}Fe_{19}$ 1.8-5 (wedge)/MgO 2 | 0.04 ±0.01 | -1.43 ±0.07 | Flip-chip FMR |
| C1 | $Si/SiO_2$/Ta 1/Pt 4/$Fe_{60}Co_{20}B_{20}$ 1.6-5.7 (wedge)/MgO 2 | 0.04 ±0.01 | -1.35 ±0.10 | ST-FMR |
| C2 | $Si/SiO_2$/Ta 1/Ta 4/$Fe_{60}Co_{20}B_{20}$ 3.0-10.7 (wedge)/MgO 2 | 0.09 ±0.01 | -2.70 ±0.08 | Flip-chip FMR |
| C3 | $Si/SiO_2$/Ta 1/MgO 2/$Fe_{60}Co_{20}B_{20}$ 1.6-5.7 (wedge)/MgO 2 | 0.20 ±0.01 | -3.26 ±0.10 | Flip-chip FMR |
| C4 | $Si/SiO_2$/Ta 1/$Au_{25}Pt_{75}$ 4/$Fe_{60}Co_{20}B_{20}$ 1.6/MgO 2 | - | - | Flip-chip FMR |
| C5 | $Si/SiO_2$/Ta 1/$Au_{25}Pt_{75}$ 4/Pt 0.5/Hf 0.25/$Fe_{60}Co_{20}B_{20}$ 1.6/MgO 2 | - | - | Flip-chip FMR |
| C6 | $Si/SiO_2$/Ta 1/$Au_{25}Pt_{75}$ 4/Pt 0.5/Hf 0.25/$Fe_{60}Co_{20}B_{20}$ 1.6/Hf 0.1/MgO 2 | - | - | Flip-chip FMR |

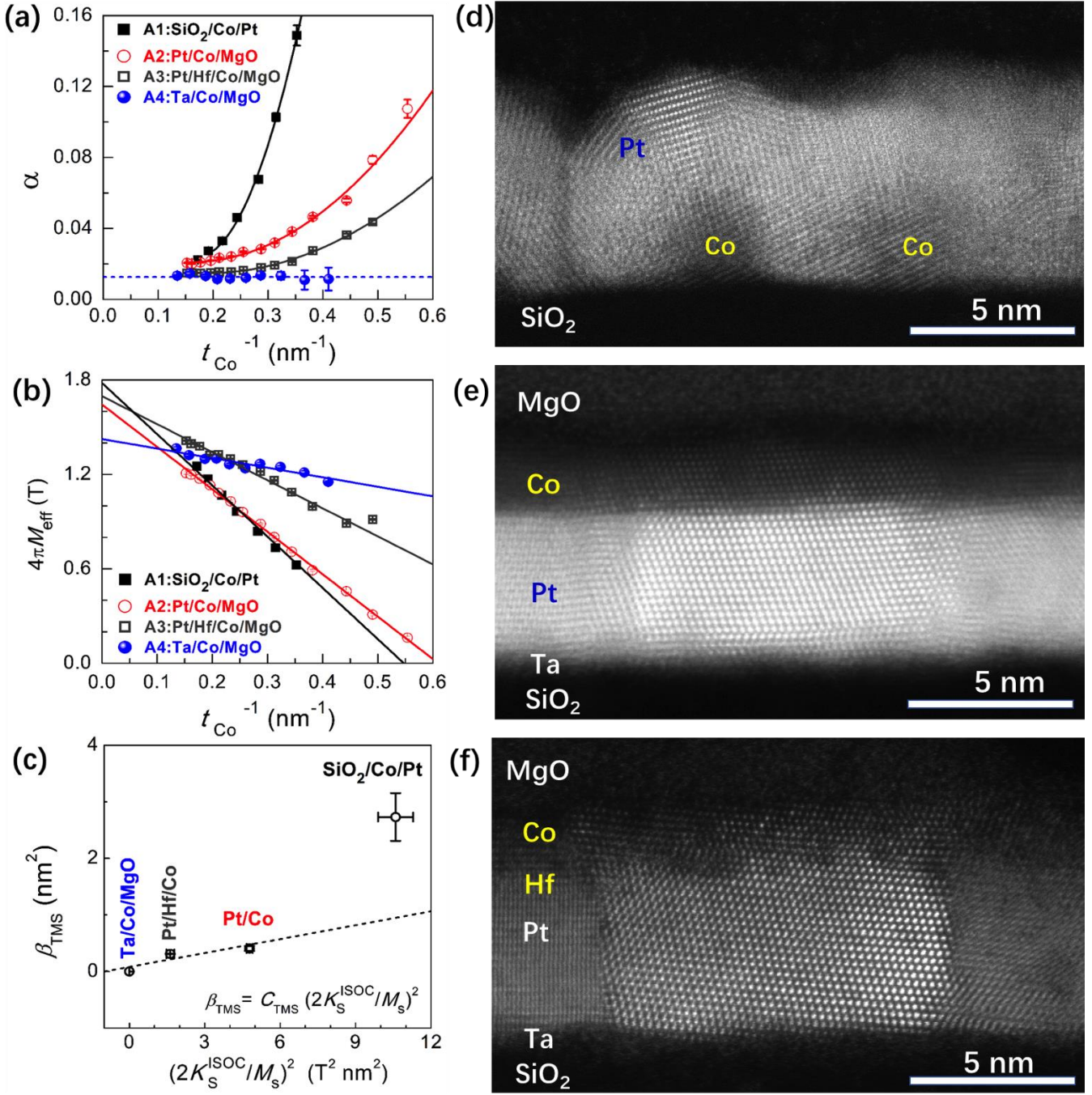

Fig. 1 Results for Co-based heterostructures. (a) $\alpha$ vs. $t_{Co}^{-1}$, (b) $4\pi M_{eff}$ vs. $t_{Co}^{-1}$, (c) $\alpha$ vs $(2K_s/M_s)^2$ for magnetic bilayers of $SiO_2/Co$ $t_{Co}/Pt$ 4.5 (series A1), Pt 4.5/Co $t_{Co}$/MgO (A2), Pt 4.5/Hf 0.3/Co $t_{Co}$/MgO (A3), and Ta 1/Co $t_{Co}$/MgO (A4). The solid lines represent the best fits of the data to Eq. (4) in (a) and to Eq. (5) in (b). Cross-sectional dark-field transmission electron microscopy images for (d) $SiO_2$/Co 2.3/Pt 4.5 (A1), (e) Pt 4.5/Co 2.5/MgO (A2), and (f) Pt 4.5/Hf 0.3/Co 2.3/MgO (A3). In all cases, both the Pt and Co layers are of polycrystalline texture, but the roughness is much greater in (d) than in (e) and (f).
7

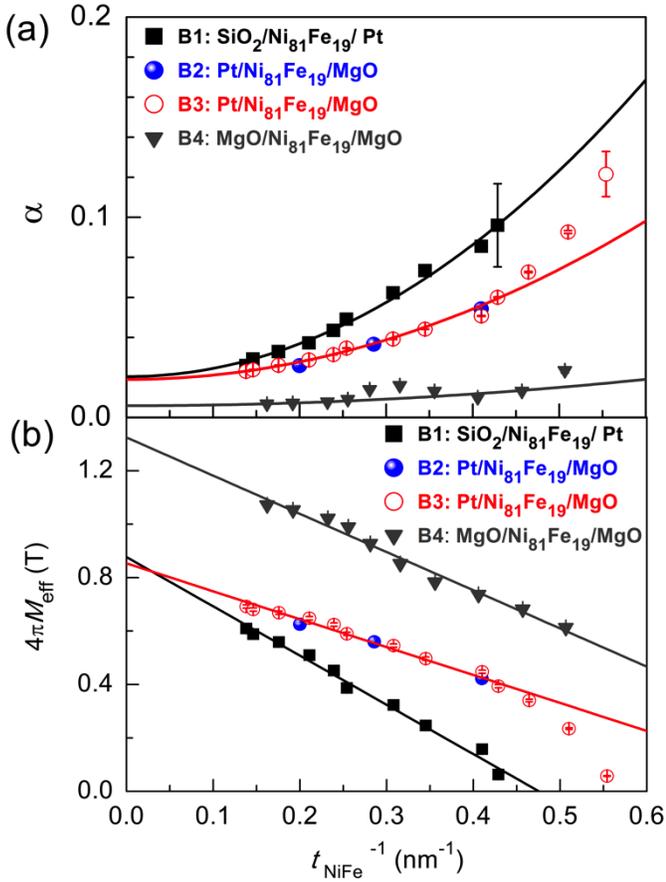

Fig. 2 Results for $Ni_{81}Fe_{19}$-based heterostructures. (a) $\alpha$ vs. $t_{NiFe}^{-1}$ and (b) $4\pi M_{eff}$ vs. $t_{NiFe}^{-1}$ for $SiO_2/Ni_{81}Fe_{19}$ $t_{NiFe}$/Pt 4.5 (series B1), Pt 4.5/$Ni_{81}Fe_{19}$ $t_{NiFe}$/MgO (B2), Pt 4.5/$Ni_{81}Fe_{19}$ 1.8, 2, 3/MgO 2 (B3), MgO 2/$Ni_{81}Fe_{19}$ wedge/MgO (B4). The lines in (a) represent the best fits of the data to Eq. (4); the straight lines in (b) represent the best linear fits of the data to Eq. (5).

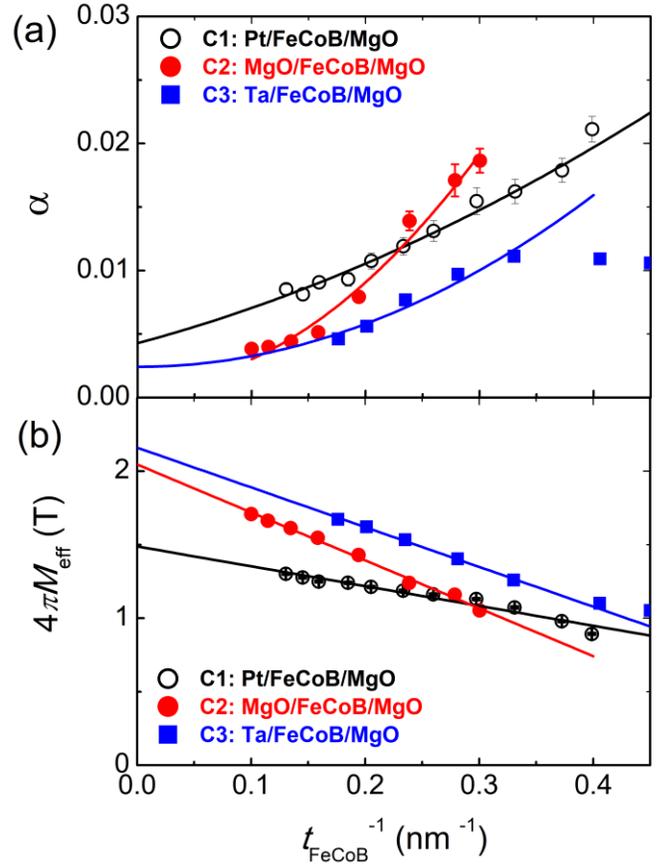

Fig. 3 Results for $Fe_{60}Co_{20}B_{20}$-based heterostructures. (a) $\alpha$ vs. $t_{FeCoB}^{-1}$ and (b) $4\pi M_{eff}$ vs. $t_{FeCoB}^{-1}$ for Pt 4/$Fe_{60}Co_{20}B_{20}$ wedge/MgO (series C1), Ta 4/$Fe_{60}Co_{20}B_{20}$ wedge/MgO (C2), and MgO 2/$Fe_{60}Co_{20}B_{20}$ wedge/MgO 2 (C3). The lines in (a) represent the best fits of the data to Eq. (4); the straight lines in (b) represent the best linear fits of the data to Eq. (5).

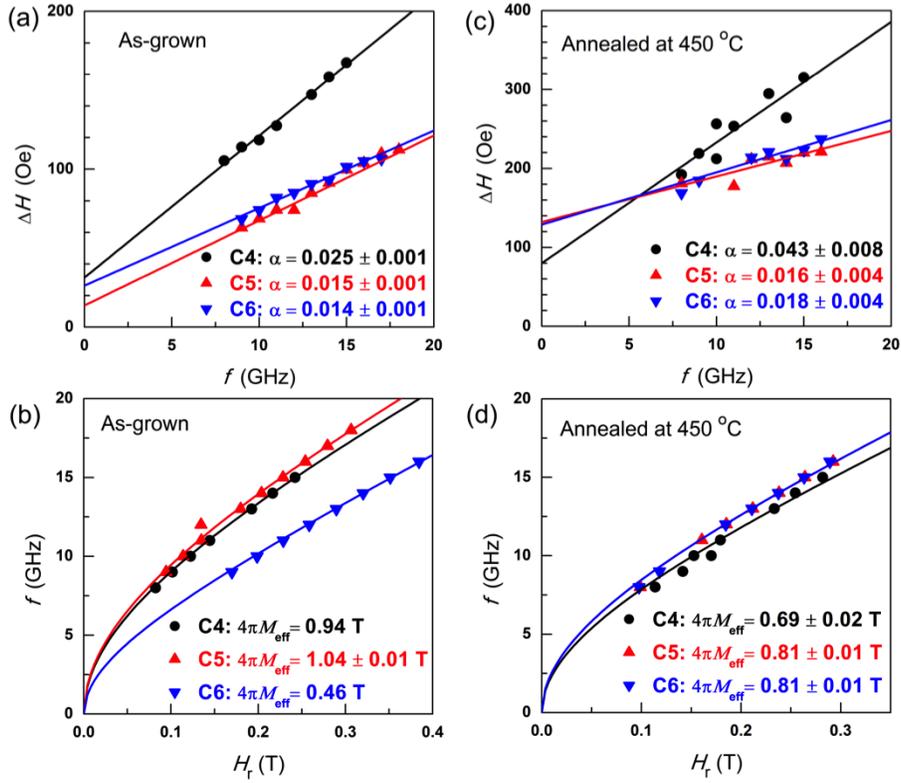

Fig. 4 Effects of TMS in $Au_{25}Pt_{75}/Fe_{60}Co_{20}B_{20}/MgO$ heterostructures. (a) The dependence of the ferromagnetic resonance (FMR) linewidth ($\Delta H$) on frequency ($f$) and (b) $f$ vs. the FMR resonance field ($H_r$) for $Au_{25}Pt_{75}$ 4/$Fe_{60}Co_{20}B_{20}$ 1.6/MgO (sample C4), $Au_{25}Pt_{75}$ 4/Pt 0.5/Hf 0.25/$Fe_{60}Co_{20}B_{20}$ 1.6/MgO (C5), and $Au_{25}Pt_{75}$ 4/Pt 0.5/Hf 0.25/$Fe_{60}Co_{20}B_{20}$ 1.6/Hf 0.1/MgO (C6). (c) and (d) The same quantities after annealing the samples at 450 C for 1 hour. All of the data were taken with flip-chip FMR on unpatterned sample pieces. The lines represent the best fits of the data to Eq. (2) in (a) and (c) and to Eq. (3) in (b) and (d).

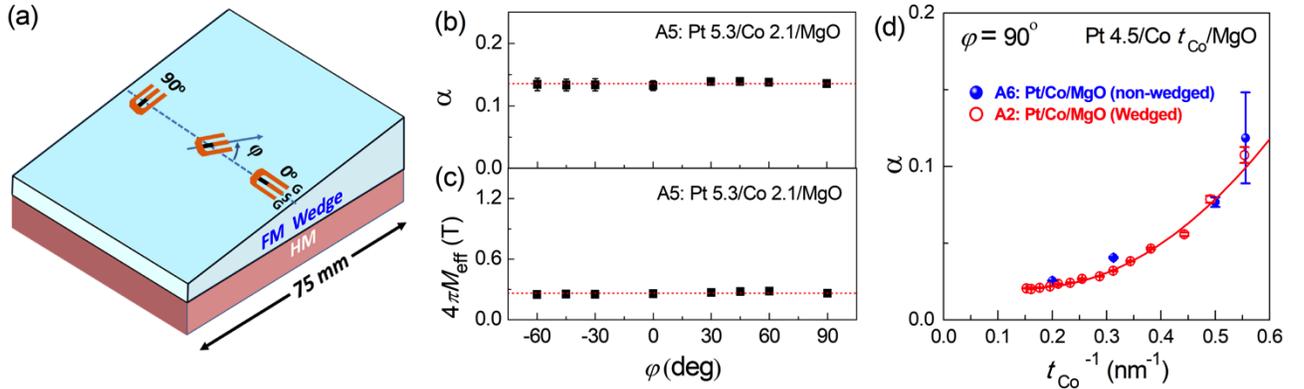

Fig. 5. (a) Schematic depiction of ST-FMR devices with different orientations ($\varphi$) with respect to the gradient of a FM wedge layer. The dependence of (b) magnetic damping and (c) demagnetization field on $\varphi$ for Pt 5.3/Co 2.1 bilayers from a Co-wedged sample, indicating the absence of any anisotropy due to the wedge. (d) Dependence of the magnetic damping on $t_{Co}^{-1}$ for Pt 4.5/Co 1.8-6.6/MgO where $t_{Co}$ was varied either by using a Co wedge (series A2) or using separate non-wedged Co layers (A6). All the data were taken using ST-FMR.

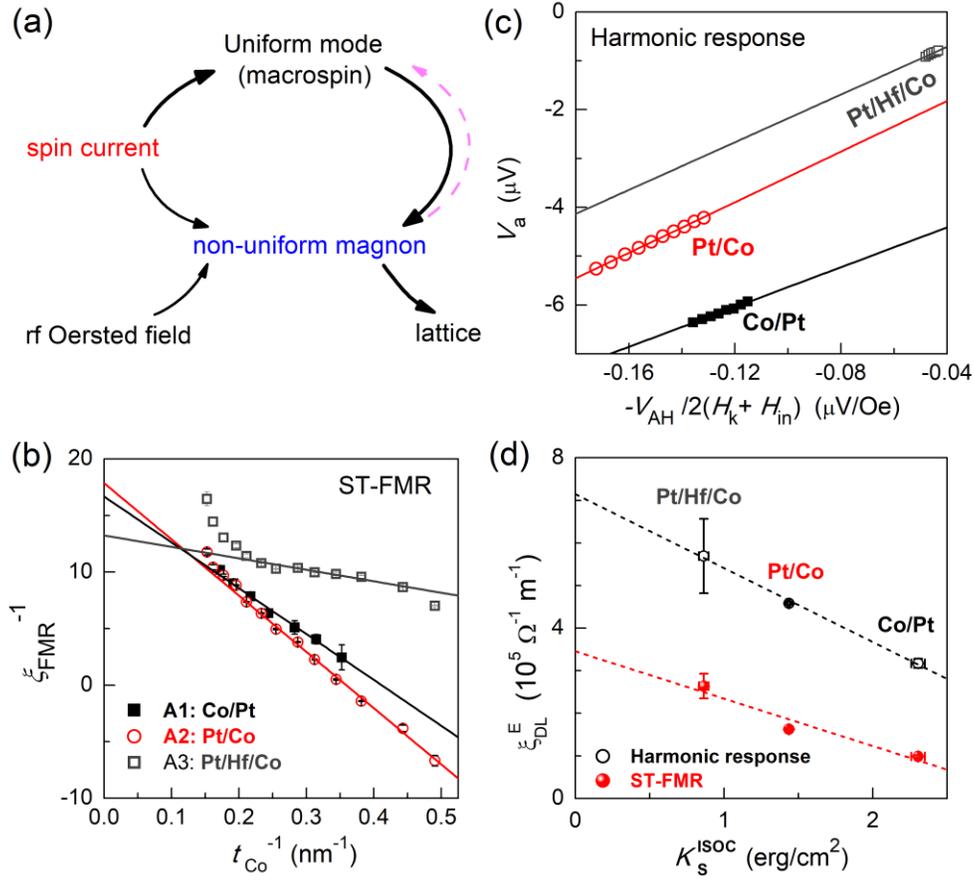

Fig. 6. (a) Possible involvement of the non-uniform magnons in spin-orbit torque process. (b) $\xi_{FMR}^{-1}$ vs $t_{Co}^{-1}$ (ST-FMR measurement), (c) Linear dependence of $V_a$ on $-V_{AH}/2(H_k + H_{in})$ (Harmonic response, $t_{Co}$ = 2.3 or 2.5 nm), (d) $\xi_{DL}^{E}$ vs $K_s^{ISOC}$ of Co/Pt (sample A1), Pt/Co (sample A2) and Pt/Hf/Co (sample A3). The Pt layer is 4.5 nm thick in all cases and the Hf layer is 0.3 nm thick. The straight lines in (b)-(d) represent the best linear fits of the data. In (d), $\xi_{DL}^{E}$ for Pt/Hf/Co samples is the value after the correction for the (28 ± 6)% attenuation of spin current in the 0.3 nm thick Hf spacer layer (see ref. [39] for more details).